\documentclass[12pt]{article}
\usepackage{amsmath}
\usepackage{amssymb}
\usepackage{amsfonts}
\usepackage{color}

\title{Rational extensions of the Dunkl oscillator in the plane and exceptional orthogonal polynomials}
\author{C.\ Quesne \\
{\small\sl Physique Nucl\'eaire Th\'eorique et Physique Math\'ematique, 
Universit\'e Libre de Bruxelles,} \\ 
{\small \sl Campus de la Plaine CP229, Boulevard~du Triomphe, B-1050
Brussels, Belgium} \\
{\small \sl Christiane.Quesne@ulb.be}}
\date{ }
\begin{document}
\baselineskip=22pt plus 1pt minus 1pt
\maketitle

\begin{abstract}
It is shown that rational extensions of the isotropic Dunkl oscillator in the plane can be obtained by adding some terms either to the radial equation or to the angular one obtained in the polar coordinates approach. In the former case, the isotropic harmonic oscillator is replaced by an isotropic anharmonic one, whose wavefunctions are expressed in terms of $X_m$-Laguerre exceptional orthogonal polynomials. In the latter, it becomes an anisotropic potential, whose explicit form has been found in the simplest case associated with $X_1$-Jacobi exceptional orthogonal polynomials.
\end{abstract}

\vspace{0.5cm}

\noindent
{\sl PACS}: 02.30.Gp; 03.65.Fd; 03.65.Ge 

\noindent
{\sl Keywords}: Dunkl quantum mechanics; reflection operators; exceptional orthogonal polynomials
 
\newpage
%
%
\section{Introduction}

In 1950, Wigner \cite{wigner} introduced the use of the reflection operator in quantum mechanics and soon after Yang applied it to the harmonic oscillator \cite{yang}. Dunkl independently considered sets of differential-difference operators associated with finite reflection groups \cite{dunkl89} and now referred to as Dunkl operators. Such operators have been found very useful in mathematics \cite{dunkl14}, as well as in physics, where they have been applied, for instance, for bosonizing supersymmetric quantum mechanics \cite{plyu} or some generalizations thereof (see, e.~g., Ref.~\cite{cq21} and references quoted therein), for building an exchange operator formalism in Calogero-Sutherland-Moser type models \cite{lapointe, cq95}, and for proving the superintegrability of some models~\cite{cq10}.\par
%
%
During recent years, there has been much interest in studying exactly solvable models in a deformed quantum mechanics (sometimes called Wigner-Dunkl quantum mechanics), wherein ordinary derivatives are replaced by Dunkl ones. One may quote, for instance, several variants of Dunkl oscillators \cite{genest13a, genest14a, genest13b, genest14b} and Dunkl-Coulomb problems \cite{genest15, ghazou}.\par
%
In a recent work \cite{cq23}, a new extension of such models was proposed wherein the replacement of ordinary derivatives by Dunkl ones was combined with that of classical orthogonal polynomials (COPs) by exceptional orthogonal polynomials (EOPs). The latter, which form orthogonal and complete polynomial sets although they admit some gaps in the sequence of their degrees in contrast with the former \cite{gomez09}, made their appearance in standard quantum mechanics in relation to Darboux transformations and shape invariant potentials \cite{cq08, cq09}.They have been used there to build infinite families of shape invariant potentials connected with $X_m$ EOPs \cite{odake09} or with multi-indexed families of $X_{m_1 m_2 \ldots m_k}$ EOPs \cite{gomez12, odake11}.\par
%
%
The purpose of Ref.~\cite{cq23} was to show that the extensions of the exactly solvable quantum mechanical problems connected with the replacement of ordinary derivatives by Dunkl ones and with that of COPs by EOPs could be easily combined. To this aim, the simplest example of the Dunkl oscillator on the line was considered and three different types of rational extensions were constructed in connection with the three infinite families of $X_m$-Laguerre EOPs \cite{cq09, grandati, cq11, liaw}.\par
%
%
In the present paper, we analyze the more elaborate problem of the (isotropic) Dunkl oscillator in the plane. Starting from its solution in cartesian coordinates and combining the results of Ref.~\cite{cq23} in both coordinates would have been rather trivial. Instead of this, we propose here an innovative approach based on its solution in polar coordinates and we show that several rational extensions can be constructed by adding some terms either to the radial equation or to the angular one.\par
%
%
\section{The Dunkl Oscillator in the Plane in Polar Coordinates}

The isotropic Dunkl oscillator in the plane is defined by the Hamiltonian \cite{genest13a}
\begin{equation}
  H = \tfrac{1}{2} \left(- D_1^2 - D_2^2 + x_1^2 + x_2^2\right), 
\end{equation}
where $D_i$, $i=1$, 2, denotes a Dunkl derivative
\begin{equation}
  D_i = \partial_{x_i} + \frac{\mu_i}{x_i} (1-R_i), \qquad \mu_i > - \frac{1}{2},
\end{equation}
and $R_i$ is the reflection operator defined by $R_i f(x_i) = f(-x_i)$. Hence $H$ can be written as
\begin{equation}
  H = \frac{1}{2} \left(-\partial_{x_1}^2 - \partial_{x_2}^2 - \frac{2\mu_1}{x_1}\partial_{x_1} - \frac{2\mu_2}
  {x_2} \partial_{x_2} + \frac{\mu_1}{x_1^2}(1-R_1) + \frac{\mu_2}{x_2^2}(1-R_2) + x_1^2 + x_2^2\right). 
\end{equation}
\par
%
%
In polar coordinates $x_1=\rho \cos \phi$, $x_2 = \rho \sin \phi$ (with $0 < \rho < +\infty$, $0 < \phi < 2\pi$), $H$ becomes
\begin{equation}
  H = A_{\rho} + \frac{1}{\rho^2} B_{\phi},  \label{eq:H}
\end{equation}
where
\begin{equation}
\begin{split}
  A_{\rho} &= \frac{1}{2}\left(-\partial_{\rho}^2 - \frac{2\mu_1 + 2\mu_2 + 1}{\rho} \partial_{\rho}
       + \rho^2\right), \\
  B_{\phi} & = \frac{1}{2}\left(- \partial_{\phi}^2 + 2(\mu_1\tan\phi - \mu_2\cot\phi)\partial_{\phi}
       + \frac{\mu_1(1-R_1)}{\cos^2\phi} + \frac{\mu_2(1-R_2)}{\sin^2\phi}\right),  
\end{split} \label{eq:DO-polar}
\end{equation}
and the action of the reflection operators is now given by
\begin{equation}
 R_1 f(\rho,\phi) = f(\rho, \pi-\phi), \qquad R_2 f(\rho, \phi) = f(\rho, -\phi).
\end{equation}
The corresponding Schr\"odinger equation $H \Psi(\rho,\phi) = {\cal E} \Psi(\rho, \phi)$ is separable and on setting $\Psi(\rho,\phi) = R(\rho) \Phi(\phi)$, one finds the pair of equations
\begin{align}
  &\left(A_{\rho} + \frac{M^2}{2\rho^2}\right) R(\rho) = {\cal E} R(\rho), \label{eq:rad-eqn} \\
  &B_{\phi} \Phi(\phi) = \frac{M^2}{2} \Phi(\phi),  \label{eq:ang-eqn}
\end{align}
where $M^2/2$ is the separation constant.\par
%
%
Since $H$ commutes with $R_1$ and $R_2$, its eigenstates may be labelled by the eigenvalues $s_1$, $s_2=\pm 1$ of $R_1$ and $R_2$. It is actually convenient to set $s_i=1-2\epsilon_i$, where $\epsilon_i=0$ for $s_i=+1$ and $\epsilon_i=1$ for $s_i=-1$. Then $\Phi(\phi)$ will be denoted by $\Phi^{(\epsilon_1,\epsilon_2)}(\phi)$, which is a solution of the equation
\begin{align}
  &\left(-\frac{d^2}{d\phi^2} + 2(\mu_1\tan\phi - \mu_2\cot\phi)\frac{d}{d\phi} + 2\mu_1\epsilon_1\sec^2\phi
      + 2\mu_2\epsilon_2\csc^2\phi - M^2\right) \Phi^{(\epsilon_1,\epsilon_2)}(\phi) \nonumber \\
  &\quad = 0.  \label{eq:ang-eqn-bis}
\end{align}
The change of function $\Phi^{(\epsilon_1,\epsilon_2)}(\phi) = (\cos\phi)^{-\mu_1} (\sin\phi)^{-\mu_2} \Xi^{(\epsilon_1,\epsilon_2)}(\phi)$ transforms this equation into
\begin{align}
  &\left(-\frac{d^2}{d\phi^2} + \mu_1(\mu_1-1+2\epsilon_1)\sec^2\phi + \mu_2(\mu_2-1+2\epsilon_2)
      \csc^2\phi - (\mu_1+\mu_2)^2 - M^2\right) \Xi^{(\epsilon_1,\epsilon_2)}(\phi) \nonumber \\
  &\quad = 0.  \label{eq:xi-eqn}
\end{align}
\par
%
%
Such an equation looks like the well-known Schr\"odinger equation for the P\"oschl-Teller I (or PT I) potential $V_{A,B}(x) = A(A-1)\sec^2 x + B(B-1) \csc^2 x$ \textcolor{red}{\cite{poschl, dabrowska}}, except that in the latter $0<x<\pi/2$ while $0<\phi<2\pi$ in (\ref{eq:xi-eqn}).\footnote{\label{footnote}Note that some results for the Scarf I potential $\bar{V}_{\bar{A},\bar{B}}(\bar{x})= [\bar{A}(\bar{A}-1)+\bar{B}^2]\sec^2\bar{x} - \bar{B}(2\bar{A}-1)\sec\bar{x}\tan\bar{x}$, $-\frac{\pi}{2}<\bar{x}<\frac{\pi}{2}$, $0<\bar{B}<\bar{A}-1$, may alternatively be used since it is related to the PT I potential by the changes of parameters and of variable $A=\bar{A}-\bar{B}$, $B=\bar{A}+\bar{B}$, $x = \frac{1}{2}\left(\bar{x}+\frac{\pi}{2}\right)$.} This means that from the PT I wavefunctions
\begin{equation}
  \psi^{(A,B)}_{\nu}(x) = \left(\frac{2(A+B+2\nu)\nu! \Gamma(A+B+\nu)}{\Gamma(A+\nu+\frac{1}{2})
  \Gamma(B+\nu+\frac{1}{2})}\right)^{1/2} \cos^A x \sin^B x P^{(A-\frac{1}{2},B-\frac{1}{2})}_{\nu}(-\cos2x),
\end{equation}
written in terms of Jacobi polynomials and with eigenvalues $E_{\nu}(A,B) = (A+B+2\nu)^2$, $\nu=0$, 1, 2,~\ldots, one can get the solutions $\Xi^{(\epsilon_1,\epsilon_2)}_n(\phi)$ of (\ref{eq:ang-eqn-bis}) by performing the replacements $x\to\phi$, $A\to\mu_1+\epsilon_1$, $B\to\mu_2+\epsilon_2$, $\nu=n-\frac{1}{2}(\epsilon_1 +
\epsilon_2)$, and multiplying the result by an extra factor 1/2, which takes the change of normalization into account. This leads to the result
\begin{align}
  \Phi^{(\epsilon_1,\epsilon_2)}_n(\phi) &= \left(\frac{(2n+\mu_1+\mu_2)\left(n-\frac{\epsilon_1+\epsilon_2}
     {2}\right)! \Gamma\left(n+\mu_1+\mu_2+\frac{\epsilon_1+\epsilon_2}{2}\right)}{2\Gamma\left(n+\mu_1+
     \frac{1+\epsilon_1-\epsilon_2}{2}\right)\Gamma\left(n+\mu_2+\frac{1+\epsilon_2-\epsilon_1}{2}\right)} 
     \right)^{1/2} \nonumber \\
  &\quad \times \cos^{\epsilon_1}\phi \sin^{\epsilon_2}\phi P^{\left(\mu_1+\epsilon_1-\frac{1}{2},\mu_2+
     \epsilon_2-\frac{1}{2}\right)}_{n-(\epsilon_1+\epsilon_2)/2}(-\cos2\phi),  \label{eq:Phi} 
\end{align}
corresponding to
\begin{equation}
  M^2 = 4n(n+\mu_1+\mu_2).  \label{eq:M}
\end{equation}
In (\ref{eq:Phi}) and (\ref{eq:M}), $n$ runs over all nonnegative integers for $\epsilon_1=\epsilon_2=0$, all positive integers for $\epsilon_1=\epsilon_2=1$, and all positive half-integers for $\epsilon_1=0$, $\epsilon_2=1$ or $\epsilon_1=1$, $\epsilon_2=0$.\par
%
%
With $M^2$ given in (\ref{eq:M}) and the change of function $R(\rho) = \rho^{-\mu_1-\mu_2-\frac{1}{2}} Q(\rho)$, the radial differential equation (\ref{eq:rad-eqn}) is changed into
\begin{equation}
  \left(-\frac{d^2}{d\rho^2} + \frac{\left(2n+\mu_1+\mu_2-\frac{1}{2} \right)\left(2n+\mu_1+\mu_2
  +\frac{1}{2}\right)}{\rho^2} + \rho^2\right) Q(\rho) = 2 {\cal E} Q(\rho),
\end{equation}
which is similar to that of the three-dimensional oscillator $V_l(x) = \frac{l(l+1)}{x^2}+ \frac{1}{4}\omega^2 x^2$,  $0<x<+\infty$. From the known solutions of the latter in terms of Laguerre polynomials (see, e.g., Eq.~(2.3) of Ref.~\cite{cq09}) and the replacements $x \to \rho$, $\omega \to 2$, $l \to 2n+\mu_1+\mu_2-\frac{1}{2}$, $\nu \to k$, one therefore directly gets
\begin{equation}
  R_{k,n}(\rho) = \left(\frac{2 k!}{\Gamma(k+2n+\mu_1+\mu_2+1)}\right)^{1/2} \rho^{2n} e^{-\frac{1}{2}
  \rho^2} L^{(2n+\mu_1+\mu_2)}_k(\rho^2), \label{eq:R}
\end{equation}
corresponding to
\begin{equation}
  {\cal E}_{k,n} = 2k + 2n + \mu_1 + \mu_2 + 1, \qquad k=0, 1, 2, \ldots.  \label{eq:E}
\end{equation}
\par
%
%
Note that the angular wavefunctions (\ref{eq:Phi}) satisfy the orthonormality condition
\begin{equation}
  \int_0^{2\pi} \Phi^{(\epsilon'_1, \epsilon'_2)}_{n'}(\phi) \Phi^{(\epsilon_1,\epsilon_2)}_n(\phi)
  |\cos\phi|^{2\mu_1} |\sin\phi|^{2\mu_2} d\phi = \delta_{n',n} \delta_{\epsilon'_1,\epsilon_1}
  \delta_{\epsilon'_2,\epsilon_2},  \label{eq:ortho-ang}
\end{equation}
while the radial wavefunctions (\ref{eq:R}) are such that
\begin{equation}
  \int_0^{+\infty} R_{k',n}(\rho) R_{k,n}(\rho) \rho^{2\mu_1+2\mu_2+1} d\rho = \delta_{k',k}.
  \label{eq:ortho-rad}
\end{equation}
\par
%
%
\section{Rationally Extending the Radial Equation}

The relation of the radial equation (\ref{eq:rad-eqn}) to the radial equation of the three-dimensional oscillator makes it easy to extend the former by starting from the known rational extensions of the latter. In the simplest case connected with the three different types of $X_m$-Laguerre EOPs (see, e.g., Refs.~\cite{cq23, cq09, grandati, cq11, liaw}), one therefore replaces the harmonic oscillator potential $\frac{1}{2}\rho^2$ in (\ref{eq:DO-polar}) by
\begin{equation}
  \frac{1}{2}\rho^2 - 2\left\{\frac{\dot{g}^{(\alpha)}_m}{g^{(\alpha)}_m} + 2\rho^2 \left[
  \frac{\ddot{g}^{(\alpha)}_m}{g^{(\alpha)}_m} - \left(\frac{\dot{g}^{(\alpha)}_m}
  {g^{(\alpha)}_m}\right)^2\right] 
  \right\},  \label{eq:HO-ext}
\end{equation}
where
\begin{equation}
  g^{(\alpha)}_m(z) = \begin{cases}
      L^{(\alpha-1)}_m(-z) & \text{for type I}, \\[0.2cm]
      L^{(-\alpha-1)}_m(z), \quad m < \alpha+1 & \text{for type II}, \\[0.2cm]
      L^{(-\alpha-1)}_m(-z), \quad m < \alpha+1, \quad \text{$m$ even} & \text{for type III},  
  \end{cases} \label{eq:g}
\end{equation}
with $\alpha = 2n+\mu_1+\mu_2$, $z=\rho^2$, and a dot denoting a derivative with respect to $z$. We shall henceforth denote $g^{(\alpha)}_m(z)$ by $g^{(\tau,\alpha)}_m(z)$, where $\tau$ is the type I, II, or III.\par
%
%
The resulting extended Hamiltonian has eigenfunctions $\Psi_{\rm ext}(\rho, \phi) = R^{(\tau,\alpha)}_{m,k,n}(\rho) \Phi^{(\epsilon_1,\epsilon_2)}_n(\phi)$, where $\Phi^{(\epsilon_1,\epsilon_2)}_n(\phi)$ remains given by (\ref{eq:Phi}), while $R^{(\tau,\alpha)}_{m,k,n}(\rho)$ can be written as
\begin{equation}
  R^{(\tau,\alpha)}_{m,k,n}(\rho) = {\cal N}^{(\tau,\alpha)}_{m,k,n} \frac{\rho^{2n} e^{-\frac{1}{2}\rho^2}}
  {g^{(\tau,\alpha)}_m(\rho^2)} L^{\tau,\alpha}_{m,k}(\rho^2)
\end{equation}
 in terms of a $k$th-degree $X_m$-Laguerre EOP of type $\tau$, where $k=m$, $m+1$, $m+2$, \ldots\ for type I or II or $k=0$, $m+1$, $m+2$, \ldots\ for type III (and, in addition, $m<2n+\mu_1+\mu_2+1$ for type II or III, as well as $m$ even for type III). The energy eigenvalues (\ref{eq:E}) are replaced by
 \begin{equation}
  {\cal E}^{(\tau,\alpha)}_{m,k,n} = 2k - 2m + 2n + \mu_1 + \mu_2 + 1,
\end{equation}
so that the spectrum remains unchanged only in type I or II case.\par
%
%
The radial wavefunctions satisfy orthonormality conditions similar to (\ref{eq:ortho-rad}) with the choices
\begin{equation}
\begin{split}
  {\cal N}^{(\rm{I},\alpha)}_{m,k,n} &= \left(\frac{2(k-m)!}{(k+2n+\mu_1+\mu_2)
        \Gamma(k+2n+\mu_1+\mu_2-m)}\right)^{1/2}, \\
  {\cal N}^{(\rm{II},\alpha)}_{m,k,n} &= \left(\frac{2(k-m)!}{(k+2n+\mu_1+\mu_2+1-2m) \Gamma(k+2n+\mu_1
        +\mu_2+2-m)} \right)^{1/2}, \\
  {\cal N}^{(\rm{III},\alpha)}_{m,k,n} &= \begin{cases}
        \left(\frac{2}{\Gamma(2n+\mu_1+\mu_2+1-m) m!}\right)^{1/2} & \text{if $k=0$}, \\
        \left(\frac{2(k-m-1)!}{k\Gamma(k+2n+\mu_1+\mu_2+1-m)}\right)^{1/2} & \text{if $k=m+1, m+2, \ldots$}.
        \end{cases}
\end{split}
\end{equation}
\par
%
%
In the $m=1$ case, for instance, the extended potential (\ref{eq:HO-ext}) reads
\begin{equation}
  \frac{1}{2} \rho^2 + \frac{2}{\rho^2+2n+\mu_1+\mu_2} - \frac{4(2n+\mu_1+\mu_2)}{(\rho^2+2n+\mu_1
  +\mu_2)^2}
\end{equation}
for type I or II.\par
%
%
\section{Rationally Extending the Angular Equation}

Extending the angular equation (\ref{eq:ang-eqn}) looks more involved because of the dependence of $B_{\phi}$ on the eigenvalues of $R_1$ and $R_2$. For this reason, we are going to restrict ourselves here to the counterpart of the simplest rational extension of the PT I potential, which reads\footnote{This extension can be directly obtained from that of the Scarf I potential given in \cite{cq08, cq09} by the changes of parameters and of variable mentioned in footnote \ref{footnote}.}
\begin{equation}
  V_{A,B,{\rm ext}}(x) = V_{A,B}(x) + \frac{8(A+B-1)}{A+B-1+(B-A)\cos 2x} - \frac{8(2A-1)(2B-1)}{[A+B-1
  +(B-A)\cos 2x]^2}
\end{equation}
and whose eigenvalues are still given by $E_{\nu}(A,B) = (A+B+2\nu)^2$, $\nu=0$, 1, 2, \ldots, while its eigenfunctions are given by
\begin{align}
  \psi^{(A,B)}_{\nu,{\rm ext}}(x) &= (B-A) \left(\frac{8(A+B+2\nu)\nu!\Gamma(A+B+\nu)}{\left(A+\nu+\frac{1}
        {2}\right)\left(B+\nu+\frac{1}{2}\right)\Gamma\left(A+\nu-\frac{1}{2}\right)
        \Gamma\left(B+\nu-\frac{1}{2}\right))}\right)^{1/2} \nonumber \\
  &\quad\times \frac{\cos^A x \sin^B x}{A+B-1+(B-A)\cos 2x} \hat{P}^{\left(A-\frac{1}{2}, B-\frac{1}{2}\right)}
        _{\nu+1}(-\cos2x)
\end{align}
in terms of $X_1$-Jacobi EOPs.\par
%
%
In the present case, Hamiltonian (\ref{eq:H}) is replaced by
\begin{equation}
  H_{\rm ext} = A_{\rho} + \frac{1}{\rho^2} B_{\phi,{\rm ext}},
\end{equation}
where $B_{\phi,{\rm ext}}$ contains some additional terms depending on the operator $\frac{1}{4}(1+R_1)(1+R_2)$, $\frac{1}{4}(1-R_1)(1-R_2)$, $\frac{1}{4}(1+R_1)(1-R_2)$, or $\frac{1}{4}(1-R_1)(1+R_2)$, selecting $(\epsilon_1,\epsilon_2) = (0,0)$, $(1,1)$, $(0,1)$, or $(1,0)$, respectively. The result reads
\begin{align}
  H_{\rm ext} &= H + \frac{1}{2}\Bigl(K_{\mu_1,\mu_2}(1+R_1)(1+R_2) + K_{\mu_1+1,\mu_2+1}(1-R_1)
       (1-R_2) \nonumber \\
  &\quad + K_{\mu_1,\mu_2+1}(1+R_1)(1-R_2) + K_{\mu_1+1,\mu_2}(1-R_1)(1+R_2)\Bigr), \label{eq:H-ext}
\end{align}
where
\begin{equation}
  K_{\mu_1,\mu_2} = \frac{1}{\rho^2}\left(\frac{\mu_1+\mu_2-1}{\mu_1+\mu_2-1
  -(\mu_1-\mu_2)\cos2\phi} - \frac{(2\mu_1-1)(2\mu_2-1)}{[\mu_1+\mu_2-1-(\mu_1-\mu_2)\cos2\phi]^2}
  \right)
\end{equation}
can be rewritten in cartesian coordinates as
\begin{align}
  K_{\mu_1,\mu_2} &= \frac{\mu_1+\mu_2-1}{(2\mu_2-1)x_1^2+(2\mu_1-1)x_2^2} - \frac{(2\mu_1-1)
        (2\mu_2-1)(x_1^2+x_2^2)}{[(2\mu_2-1)x_1^2+(2\mu_1-1)x_2^2]^2} \nonumber \\
  &= (\mu_2-\mu_1) \left(\frac{1}{(2\mu_2-1)x_1^2+(2\mu_1-1)x_2^2} - \frac{2(2\mu_1-1)x_2^2}
        {[(2\mu_2-1)x_1^2+(2\mu_1-1)x_2^2]^2}\right) \nonumber \\
  &= (\mu_1-\mu_2) \left(\frac{1}{(2\mu_2-1)x_1^2+(2\mu_1-1)x_2^2} - \frac{2(2\mu_2-1)x_1^2}
        {[(2\mu_2-1)x_1^2+(2\mu_1-1)x_2^2]^2}\right).
\end{align}
\par
%
%
Such an extended Hamiltonian still has the spectrum (\ref{eq:E}) with the radial wavefunctions given in (\ref{eq:R}), while the angular wavefunctions (\ref{eq:Phi}) are replaced by
\begin{align}
  &\Phi^{(\epsilon_1,\epsilon_2)}_{n,{\rm ext}}(\phi) = (\mu_2-\mu_1+\epsilon_2-\epsilon_1) \nonumber \\
  & \quad \times \left(\frac{2(2n+\mu_1+\mu_2)\left(n-\frac{\epsilon_1+\epsilon_2}{2}\right)!
      \Gamma\left(n+\mu_1+\mu_2+ \frac{\epsilon_1+\epsilon_2}{2}\right)}{\left(n+\mu_1+\frac{1+\epsilon_1-
      \epsilon_2}{2}\right)\left(n+\mu_2+\frac{1+\epsilon_2-\epsilon_1}{2}\right) \Gamma\left(n+\mu_1
      + \frac{\epsilon_1-\epsilon_2-1}{2}\right) \Gamma\left(n+\mu_2+\frac{\epsilon_2-\epsilon_1-1}{2}\right)}
      \right)^{1/2} \nonumber \\
  &\quad \times [\mu_1+\mu_2+\epsilon_1+\epsilon_2-1+(\mu_2-\mu_1+\epsilon_2-\epsilon_1)\cos2\phi]
      ^{-1} \nonumber \\
  &\quad \times \hat{P}^{\left(\mu_1-\frac{1}{2}+\epsilon_1,\mu_2-\frac{1}{2}+\epsilon_2\right)}_{n+1
      -(\epsilon_1+\epsilon_2)/2}(-\cos2\phi), 
\end{align}
where $n$ takes the same values as in Sec.~2 and the orthonormality condition remains given by an equation similar to (\ref{eq:ortho-ang}).\par
%
%
The operator $H_{\rm ext}$ defined in (\ref{eq:H-ext}) may alternatively be written as
\begin{equation}
  H_{\rm ext} = \frac{1}{2} \Bigl(- D_1^2 - D_2^2 + x_1^2 + x_2^2 + L^{(0,0)}_{\mu_1,\mu_2} +
  L^{(1,0)}_{\mu_1,\mu_2} R_1 + L^{(0,1)}_{\mu_1,\mu_2} R_2 + L^{(1,1)}_{\mu_1,\mu_2} R_1 R_2\Bigr),
\end{equation}
where
\begin{equation}
  L^{(p,q)}_{\mu_1,\mu_2} = K_{\mu_1,\mu_2} + (-1)^p K_{\mu_1+1,\mu_2} + (-1)^q K_{\mu_1,\mu_2+1}
  + (-1)^{p+q} K_{\mu_1+1,\mu_2+1},
\end{equation}
with $p, q \in \{0,1\}$.\par
%
%
It is also possible to express $H_{\rm ext}$ in terms of some extended Dunkl derivatives
\begin{equation}
  \hat{D}_i = D_i + F_i R_i = \partial_{x_i} + \frac{\mu_i}{x_i}(1-R_i) + F_i R_i, \qquad i=1, 2,
\end{equation}
where
\begin{equation}
  F_i = F_i^{(\mu_1,\mu_2)} + (-1)^i F_i^{(\mu_1+1,\mu_2)} - (-1)^i F_i^{(\mu_1,\mu_2+1)} - F_i^{(\mu_1+1,
  \mu_2+1)}, \qquad i=1,2,
\end{equation}
with
\begin{equation}
  F_1^{(\mu_1,\mu_2)} = \frac{(\mu_2-\mu_1)x_1}{(2\mu_2-1)x_1^2+(2\mu_1-1)x_2^2}, \qquad
  F_2^{(\mu_1,\mu_2)} = \frac{(\mu_1-\mu_2)x_2}{(2\mu_2-1)x_1^2+(2\mu_1-1)x_2^2}.
\end{equation}
Then it takes the form
\begin{equation}
\begin{split}
  &H_{\rm ext} = \frac{1}{2} \Bigl(- \hat{D}_1^2 - \hat{D}_2^2 + x_1^2 + x_2^2 + G_1 + G_2 R_1 R_2\Bigr), \\
  &G_1 = L^{(0,0)}_{\mu_1,\mu_2} + \frac{2\mu_1}{x_1}F_1 - F_1^2 + \frac{2\mu_2}{x_2}F_2 - F_2^2, \\
  &G_2 = L^{(1,1)}_{\mu_1,\mu_2}.
\end{split}
\end{equation}
\par
%
%
\section{Conclusion}

In this work, we have shown that some rational extensions of the Dunkl oscillator in the plane can be obtained by adding some terms either to the radial equation or to the angular one, obtained in the polar coordinates approach. In the former case, the isotropic harmonic oscillator is replaced by an isotropic anharmonic one, whose wavefunctions are expressed in terms of $X_m$-Laguerre EOPs. In the latter, it becomes an anisotropic potential, whose explicit form has been found in the simplest case associated with $X_1$-Jacobi EOPs.\par
%
%
Generalizing the present approach to potentials related to other EOPs would be an interesting problem for future study.\par
%
%
\section*{Acknowledgements}

This work was supported by the Fonds de la Recherche Scientifique--FNRS under Grant No.\ 4.45.10.08.\par
%
%

\end{document}